\documentclass{emulateapj}
\usepackage{amsmath}

\shorttitle{Emission due to the Kilonova Ejecta in GW 170817}
\shortauthors{Asano \& To}

\begin{document}

\title{
Subsequent Nonthermal Emission due to the Kilonova Ejecta in GW 170817
}

\author{Katsuaki Asano, and Sho To}

\address{Institute for Cosmic Ray Research, The University of Tokyo,
5-1-5 Kashiwanoha, Kashiwa, Chiba 277-8582, Japan;
tosho@icrr.u-tokyo.ac.jp, asanok@icrr.u-tokyo.ac.jp}


\begin{abstract}
The ejected material at the binary neutron star merger GW 170817
was confirmed as a kilonova by UV, optical, and IR observations.
This event provides a unique opportunity to investigate the particle acceleration
at a mildly relativistic shock propagating in the circumbinary medium.
In this paper, we simulate
the nonthermal emission from electrons accelerated by the shock
induced by the kilonova ejecta with a time-dependent method.
The initial velocity and mass of the ejecta in the simulations
are obtained from the kilonova observations in GW 170817.
If the ambient density is high enough ($\geq 10^{-2}~\mbox{cm}^{-3}$),
radio, optical/IR, and X-ray
signals will be detected in a few years, though the off-axis
short gamma-ray burst models, accounting for the X-ray/radio counterpart detected
at $\sim 10$ days after the merger, implies low ambient density.
We also demonstrate that the additional low-mass ($\sim 10^{-5} M_\odot$) component
with a velocity of $0.5 c$--$0.8 c$ can reproduce the early X-ray/radio counterpart.
This alternative model allows a favorably high density to detect the nonthermal
emission due to the kilonova ejecta.
Even for a low ambient density such as $\sim 10^{-3}~\mbox{cm}^{-3}$,
depending on the microscopic parameters for the electron acceleration,
we can expect a growth of radio flux of $\sim 0.1$ mJy in a few years.
\end{abstract}

\keywords{binaries: close --- gamma-ray burst: individual (GRB 170817A) --- gravitational waves
--- radiation mechanisms: nonthermal}

\section{Introduction}
\label{sec:intro}

The binary neutron star merger detected as gravitational-wave event GW 170817
by the Advanced LIGO and Virgo \citep{abb17a} accompanied a weak short
gamma-ray burst, GRB 170817A \citep{abb17c}.
Furthermore, follow-up observations with UV, optical, and infrared telescopes
found a kilonova \citep{abb17b,arc17,cou17,cow17,sma17,tan17,val17} emitting from the mildly relativistic ejecta
\citep{lat74,ros99,hot13}
as expected in advance \citep{li98,met10,tan13}.
The ejected material will form a shock propagating in the circumbinary medium (CBM),
and electromagnetic signals
on a timescale of a few years have been predicted
\citep{nak11,pir13,ros13,tak14,hot15,hot16}.
Electrons are accelerated at the shock and emit nonthermal
synchrotron photons from the radio to X-ray range, which is the mildly relativistic
counterpart to the emission from the supernova remnant (nonrelativistic)
or the gamma-ray burst (GRB) afterglow (ultrarelativistic).
Hereafter we call this possible phenomenon ``kilonova afterglow,''
while the previous studies focused mainly on radio emission.

GW 170817 is well localized, and its distance \citep[$\sim 40$ Mpc;][]{im17}
is exceptionally close. The physical properties of the ejecta
are also well constrained by the kilonova observations.
Therefore, this is a golden opportunity
to verify whether electrons are efficiently accelerated even in
mildly relativistic cases as seen in ultrarelativistic shocks
of GRB afterglows.
In the late phase of the GRB afterglow, the shock speed may be mildly relativistic.
However, the radio emission at this stage may be dominated by the emission
from the remnant electrons
accelerated at the relativistic stage.

The flux of the kilonova afterglow largely depends on the CBM density $n$.
In the context of the off-axis GRB afterglow model,
the X-ray and radio counterparts of GW 170817 at $\sim 10$ days after the merger
\citep{ale17,mar17,tro17} imply a low CBM density of $n \leq 10^{-3}~\mbox{cm}^{-3}$
\citep[see also][]{iok17}.
In those models, the late detection of the afterglow
is explained by the expansion of the beamed emission cone
due to the deceleration.
Since a highly relativistic ejecta as implied for typical short GRBs
rapidly decelerates, a too-high CBM density results in a too-early onset
of the off-axis afterglow.
The low density in the off-axis GRB models
leads to a very dim flux of the emission from the shocked CBM.

On the other hand, alternative models for the X-ray and radio counterparts
have been proposed \citep[e.g.][]{bro17,got17,kas17,murg17}.
In those models, the emission is due to the mildly relativistic
ejecta such as a cocoon or a wide jet.
The subsequent rising of the radio and X-ray fluxes as far as $\sim 100$ days
\citep{moo17,rua17} further supports the mildly relativistic scenario, in which
the onset time of the radio and X-ray emission
can agree with the deceleration time of the ejecta by adjusting the
initial velocity even with a high CBM density.
Although a model with a highly relativistic jet can still reconcile
the rising light curves \citep{laz17}, a very low CBM density
($\sim 10^{-5}~\mbox{cm}^{-3}$) is required to suppress the emission
from the ultrarelativistic jet.
To confirm the consistency of the off-axis GRB model,
future follow-up observations are indispensable.

In this paper, we simulate the kilonova afterglow emission
from the shocked CBM with parameter sets constrained
by the observations of GW 170817.
The evolution of the spectrum and light curves of
radio, optical/IR, and X-ray are shown.
In the most optimistic case, the radio--X-ray emission will be detected
within $\sim 1000$ days after the merger.
This is also the first demonstration of the mildly relativistic
calculation of our time-dependent numerical code in \citet{fuk17}.
Another purpose of this paper is to show significant differences
in the flux and its evolution between the simple analytic approximation
and the numerical simulation following the evolution
of the electron energy distribution.

In Section \ref{sec:form}, we shortly review
our computing method in \citet{fuk17} and show model parameters.
The radio, IR/optical, and X-ray light curves
obtained from  our code are shown in Section \ref{sec:lc}.
Only for a high CBM density case,
we expect detections of the kilonova afterglow in a few years.
However, in Section \ref{sec:earXR},
we demonstrate that the early X-ray and radio counterparts are explained
by another shock component propagating a high CBM density.
This alternative model encourages us to search for the kilonova afterglow.
The conclusions are summarized in Section \ref{sec:sum}.

\section{Model and Method}
\label{sec:form}

We adopt the time-dependent numerical code developed in \citet{fuk17}.
Our one-zone code can follow the propagation of the spherical shocked shell
even for the mildly relativistic speed with the exact shock jump condition.
The electron and photon energy distributions in the shell
are also calculated taking into account the injection of the nonthermal electrons,
radiative cooling, adiabatic cooling, synchrotron self-absorption,
and photon escape.
The energy and arrival time of photons escaped from the entire shell surface
are consistently transformed into those for an observer
with the effects of the curvature and relativistic motion of the shell.

We adopt the conventional form of
the electron spectrum at the injection: the single power law
with a minimum Lorentz factor $\gamma_{\rm m}$ and high-energy exponential cutoff.
Given the shock speed, $\gamma_{\rm m}$ is calculated
with the exact jump condition
and standard microscopic parameters for the nonthermal electrons:
the energy fraction $\epsilon_{\rm e}$,
number fraction $\eta$, and power law index $p$
\citep[see][for details]{fuk17}.
The evolution of the magnetic field is obtained with the parameter $\epsilon_B$,
which is the energy fraction of the magnetic field to the dissipated energy
at the shock.
The electron maximum energy
is calculated considering the balance of the acceleration and radiative cooling
with the acceleration time scale $20 c r_{\rm L}/(3 v_{\rm sh}^2)$,
where $v_{\rm sh}$ is the velocity of the shock front in the CBM frame,
and $r_{\rm L}$ is the Larmor radius of the particle..

As shown in \citet{fuk17}, the exact evolutions of the shock speed
and electron energy distribution lead to earlier peak time of the flux
than the analytical formula.
The adiabatic cooling significantly
affects the electron energy distribution and resultant photon spectrum.
The spectral peak flux at the cooling frequency is suppressed compared
to the flux obtained with the broken power law approximation
\citep[see also][]{pet09,pen14,uhm14}.
The analytical approximation may be optimistic to discuss
the detectability of the kilonova afterglow.
To discuss the uncertainty of the parameters from observation,
the comparison with the numerical results is useful.

The mass and velocity of the ejecta are constrained by the kilonova observations.
Here, we refer to the two-component model in \citet{cow17}:
fast component (mass $0.01 M_{\odot}$ and velocity $v=0.27 c$)
and slow component ($0.04 M_{\odot}$ and $v=0.12 c$).
The model implies a total mass of $0.05 M_{\odot}$
and kinetic energy of $1.2 \times 10^{51}$ erg,
which seems close to the highest value estimated in the numerical simulations \citep{hot13}.
If the two components are distinctly decoupled,
the slow ejecta will not catch up with the fast ejecta before the start of
its deceleration. In this case, we can neglect the contribution of the slow ejecta
to the shock dynamics in the early stage.
Hereafter, we consider two cases: (1) only the fast ejecta is taken into account
(decoupled case), and (2) the two components are well mixed so that
the average velocity is adopted as the initial velocity
for the mixed single ejecta (mixed case).
In the mixed case,
we adopt $0.16 c$ as the initial velocity for the ejecta of $0.05 M_{\odot}$.

\begin{table}[!hbtp]
	\caption{Model parameters.}
	\begin{center}
		\begin{tabular}{lcccc}
			\hline\hline
			Model & $n$ ($\mbox{cm}^{-3}$) & $\eta$ &  $\epsilon_B$ & Ejecta \\
			\hline
			A & $0.1$ & $1$ & $0.1$ & Decoupled \\
			A' & $0.1$ & $1$ & $10^{-3}$ & Decoupled \\
			B & $10^{-2}$ & $1$ & $0.1$ & Decoupled \\
			B' & $10^{-2}$ & $1$ & $0.1$ & Mixed \\
			C & $10^{-3}$ & $1$ & $0.1$ & Decoupled \\
			C' & $10^{-3}$ & $10^{-2}$ & $0.1$ & Decoupled \\
			\hline\hline
		\end{tabular}
	\end{center}
	\label{param_tab}
\end{table}

Then, the remaining model parameters are the CBM density $n$ and
the microscopic parameters $\eta$, $\varepsilon_{\rm e}$, $\varepsilon_B$,
and $p$. Here, we fix the index $p$ as 2.3, which is the typical value
in the GRB afterglow.
The parameter $\epsilon_{\rm e}$ is optimistically taken as $0.1$.
We summarize the parameters in Table \ref{param_tab}
for each model.
Basically, we focus on the cases of $\eta=1$ (all electrons are accelerated)
and $\epsilon_B=0.1$ with the decoupled ejecta assumption,
but we discuss the cases changing those parameters
(model name with prime mark) as well.

\section{Light Curves}
\label{sec:lc}

\begin{figure}[!t]
\centering
\epsscale{1.0}
\plotone{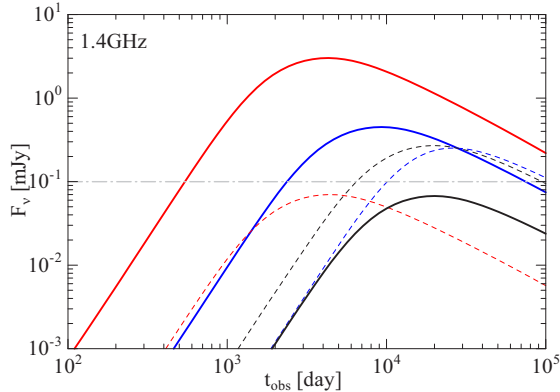}
\caption{Radio (1.4GHz) light curves for models A (red solid),
A' (red dashed), B (blue solid), B' (blue dashed), C (black solid),
and C' (black dashed).
The gray dot-dashed line shows the typical upper limit
for the 1.4GHz observations of the GRB afterglow \citep{cha12}.}
\label{frad}
\end{figure}

\begin{figure}[!t]
\centering
\epsscale{1.0}
\plotone{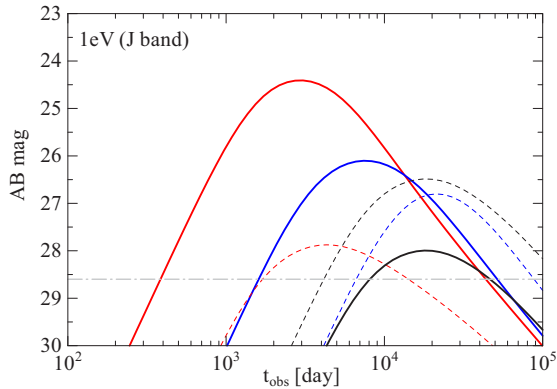}
\caption{IR (1eV, $J$ band) light curves for models A (red solid),
A' (red dashed), B (blue solid), B' (blue dashed), C (black solid),
and C' (black dashed).
The gray dot-dashed line shows the limiting magnitude
for WFC3 on {\it HST} in the 10 hr observation.}
\label{fopt}
\end{figure}

\begin{figure}[!t]
\centering
\epsscale{1.0}
\plotone{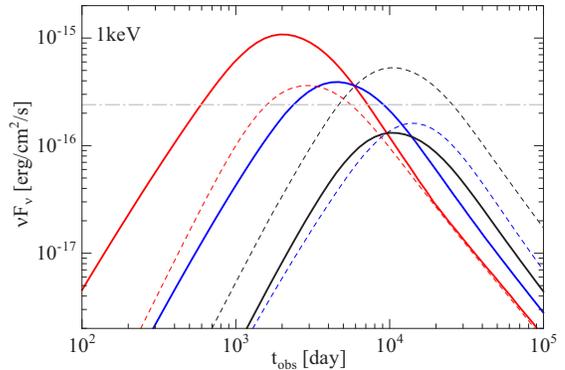}
\caption{X-ray (1keV) light curves for models A (red solid),
A' (red dashed), B (blue solid), B' (blue dashed), C (black solid),
and C' (black dashed).
The gray dot-dashed line shows the upper limit ($10^{-7}$ mJy)
for GW 170817 in the early nondetection epoch with {\it Chandra} \citep{tro17}.}
\label{fX}
\end{figure}

Figures \ref{frad}--\ref{fX} show the radio (1.4 GHz), IR ($J$ band),
and X-ray (1 keV) light curves, respectively.
The peak time roughly corresponds to the onset of the
shock deceleration.
Note that after the peak time the slow ejecta may contribute to the shock dynamics
for the decoupled models.
In such a case, the flux decay can be shallower after the peak time.
As the density decreases (see solid lines from models A to C), the peak time is delayed,
and the peak flux is suppressed \citep{nak11}.
The analytical formula in \citet{pir13} provides the peak time
\begin{eqnarray}
t_{\rm peak} &=& \frac{1}{v} \left( \frac{3 M}{4 \pi n m_{\rm p}} \right)^{1/3} \\
&\simeq& 4.0 \times 10^{3} M_{-2}^{1/3} n_{-1}^{-1/3} \beta^{-1}_{0.27}~\mbox{d},
\end{eqnarray}
where the ejecta mass $M=0.01 M_{-2} M_\odot$, $n=0.1 n_{-1}~\mbox{cm}^{-3}$,
and the initial velocity $v=0.27 \beta_{0.27} c$.
In the numerical results, the peak time slightly shifts earlier for a higher frequency.
In model A, the peak time in X-ray is $\sim 2000$ days,
while those for $J$ band and radio are $\sim 3000$ and $\sim 4000$ days, respectively.
From the analytical formula, we obtain a peak flux of
$\sim 14$ mJy at 1.4GHz for model A, which is brighter than the result in Figure \ref{frad}.
The cooling break energy at the peak time is analytically
obtained as $\sim 0.5$ eV for model A \citep{tak14}.
Then, the peak fluxes are estimated as 22.4 mag at $J$ band and $3.5 \times 10^{-15}
~\mbox{erg}~\mbox{cm}^{-2}~\mbox{s}^{-1}$ at 1 keV, respectively.

If we seriously accept the limit on the CBM density
by the off-axis GRB model \citep{iok17,mar17} based on the X-ray counterparts
at $\sim 10$ days after the merger,
only model C (and C') is within the allowed parameter space.
However, the fluxes in model C seem too dim to detect
in a few years at any wavelength in spite of its close distance and large energy.

If the density is higher than the model constraint,
the radio, IR/optical, and X-ray fluxes can be as high as
current instruments can detect as shown in Figures \ref{frad}--\ref{fX}.
Especially in the most optimistic model A,
the radio flux grows as $\sim 1$ mJy at $t \sim 10^3$ days
and $\sim 30$ $\mu$Jy at 1 yr.
The IR/optical flux becomes brighter than 26 in AB magnitude at $t \sim 10^3$ days.
The X-ray flux also reaches the detectable level
($\sim 10^{-15}~\mbox{erg}~\mbox{cm}^2~\mbox{s}^{-1}$) at $t \sim 10^3$ days.

A smaller value of $\epsilon_B$ suppresses the synchrotron flux
as shown in the light curves of model A'.
However, in the X-ray energy band, where the radiative cooling effect is significant,
the flux suppression due to low $\epsilon_B$ is not so prominent
compared to those in the other bands.
The X-ray peak flux in model A' is comparable to that
in the high-$\epsilon_B$ case (model A).

In the mixed ejecta case, the suppression of the initial velocity
extends the peak time. As a result, the flux at the early stage
is drastically suppressed (compare model B with model B' in Figures \ref{frad}--\ref{fX}),
though the total kinetic energy is larger than that in the decoupled case.
The mixed ejecta is a pessimistic assumption to detect the kilonova afterglow.

If the number fraction of the accelerated electrons is lower than unity,
the average electron energy is boosted, which implies higher emissivity.
Even for $n=10^{-3}~\mbox{cm}^{-3}$,
the model with $\eta=10^{-2}$ (model C') leads to detectable fluxes
at their peak time, $\sim 30$ yr.
Even at $\sim 10^3$ days, the radio flux can be $\sim \mu$Jy
for $\eta<1$, which is encouraging for the follow-up observation.

\begin{figure}[!t]
\centering
\epsscale{1.0}
\plotone{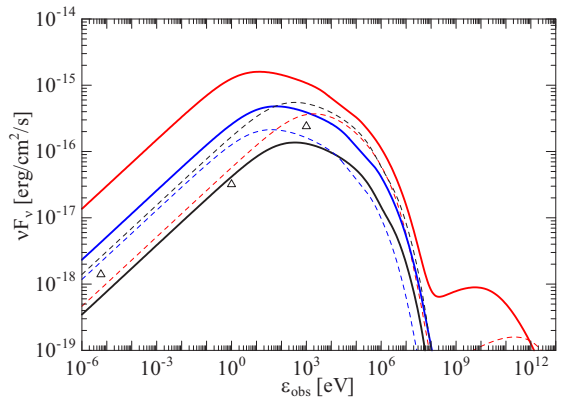}
\caption{Photon spectra at the peak time of the X-ray flux for model A (red, solid, 2,000 d),
A' (red, dashed, 3,000 d), B (blue, 4700 d), B' (blue, 13,000 d), C (black, solid, 10,000 d),
and C' (black, dashed, 10,000 d), respectively.
The detection limits depicted in Figs. \ref{frad}-\ref{fX}
are shown with triangles.}
\label{fsp}
\end{figure}

The photon spectra at the X-ray peak time are summarized
in Figure \ref{fsp}.
In the radio band, the spectrum is consistent with $F_\nu \propto \nu^{-(p-1)/2}$.
The peak energies correspond to the cooling frequency,
which is typically in UV or X-ray band.
Synchrotron self-Compton emission is expected in GeV--TeV range.
However, as shown in the spectra,
the flux level is far below the detection limit of the current instruments
({\it Fermi}-LAT and MAGIC/H.E.S.S. require flux
$>10^{-10}~\mbox{erg}~\mbox{cm}^2~\mbox{s}^{-1}$
and  $>10^{-12}~\mbox{erg}~\mbox{cm}^2~\mbox{s}^{-1}$, respectively)
even at the X-ray peak time.

\section{Spherical Model for the Early X-Ray and Radio Emission}
\label{sec:earXR}

As shown in the previous section, the low-density cases ($n \lesssim 10^{-3}~\mbox{cm}^{-3}$)
implied from the off-axis GRB models seem discouraging for detecting the kilonova afterglow.
However, the off-axis model in \citet{murg17} adopts a density of
$0.3~\mbox{cm}^{-3}$.
The spherical/wide-angle cocoon models of \citet{kas17,bro17,got17}
are alternative options to explaining the low-luminosity GRB in GW 170817.
In such models, the early X-ray and radio emissions detected at $\sim 10$ days
are not due to the decelerating relativistic jet,
and the constraint for the CBM density may be relaxed.
In this section, we assume other ejecta expanding spherically with a faster speed
than the kilonova ejecta and try to explain the early X-ray and radio emission
for a high-density case with the same numerical method.

\begin{table}[!hbtp]
	\caption{Model parameters for the early X-ray and radio emission.}
	\begin{center}
		\begin{tabular}{lcc}
			\hline\hline
			Parameter & Fast & Slow \\
			\hline
			CBM density & \multicolumn{2}{c}{$0.1~\mbox{cm}^{-3}$} \\
			$\epsilon_{\rm e}$ & \multicolumn{2}{c}{$0.05$} \\
			$\epsilon_B$ & \multicolumn{2}{c}{$0.01$} \\
			$\eta$ & \multicolumn{2}{c}{$1$} \\
			$p$ & \multicolumn{2}{c}{$2.2$} \\
			Kinetic energy & $10^{49}$~erg & $3 \times 10^{50}$~erg \\
			Ejected mass & $10^{-5} M_\odot$ & $7 \times 10^{-4} M_\odot$ \\
			Initial velocity & $0.77 c$ & $0.59 c$ \\
			\hline\hline
		\end{tabular}
	\end{center}
	\label{param_tab2}
\end{table}

\begin{figure}[!t]
\centering
\epsscale{1.0}
\plotone{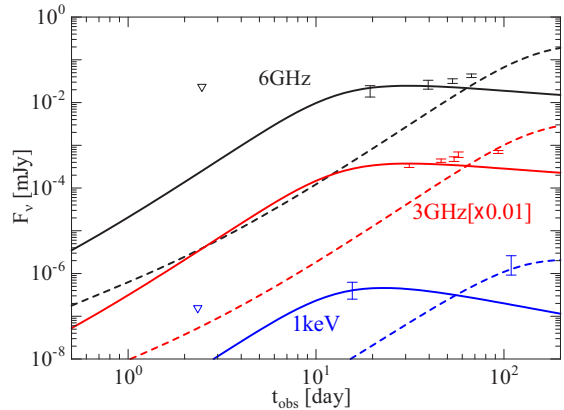}
\caption{Model light curves for 6 GHz (black), 3 GHz (red),
and X-ray (1keV; blue)
in the early phase.
The solid and dashed lines correspond to the emission
due to the fast ($v =0.77c$) and slow ($v =0.59c$) ejecta,
respectively.
See Table \ref{param_tab2} for the parameters.
The early ($<$20 days)
radio and X-ray data are taken from \citet{ale17}
and \citet{mar17}, respectively.
The late radio and X-ray data are taken from \citet{moo17}
and \citet{rua17}, respectively.}
\label{fear}
\end{figure}

First, we adopt a ``fast'' ejecta
with the initial velocity $v=0.77 c$ ($\Gamma \beta=1.2$;
see Table \ref{param_tab2} for the other parameters).
This agrees with the onsets of the X-ray and radio emission even with
a high density of $0.1~\mbox{cm}^{-3}$ 
as shown with solid lines in Figure \ref{fear}.
While \citet{murg17} also tested a similar model,
the initial velocity in our model
is significantly slower \citep[$\Gamma=5.5$ in][]{murg17}.
The energy scale, mass, and velocity are similar to those
in the cocoon model of \citet{bro17}
and the shock-heated ejecta in the model of \citet{kyu14}.
Therefore, this low-mass additional component other than the kilonova ejecta
is a reasonable assumption.

\citet{moo17} and \citet{rua17} reported the continuous rising
of the radio and X-ray light curves as far as $\sim 100$ days.
While \citet{moo17} fitted the radio data by a mildly relativistic ejecta
with a power law velocity distribution ($E(>\Gamma \beta) \propto (\Gamma \beta)^{-5}$),
our simple model with a second ``slow'' component (see Table \ref{param_tab2}
and dashed lines in Figure \ref{fear}) also agrees with the observed light curves.
\citet{laz17} claimed that an off-axis GRB model with a structured jet
can reconcile the rising light curves with a low density
($\sim 10^{-5}~\mbox{cm}^{-3}$). The future observation of the kilonova afterglow
may determine the CBM density as discussed in the previous section.

The low-mass components discussed in this section
may not contribute to the kilonova emission itself,
but they generate the delayed emission as shown in Figure \ref{fear}.
In this high-density scnario, the kilonova ejecta will catch up with
the low-mass component as it decelerates and generate the kilonova afterglow
as shown in model A in Figures \ref{frad}--\ref{fX}.
While the cocoon or wide-jet
models have structure as shown in \citet{bro17},
our simple spherical model succeeds in reproducing the light curves.
It is difficult to find signatures of the structured ejecta
from the light curves, though the structure may affect the parameter estimate.

\section{Summary}
\label{sec:sum}

GW 170817 provides a unique opportunity to verify whether
electrons are efficiently accelerated even at a mildly relativistic shock.
The early radio and X-ray emission may be due to faster ejecta than
the kilonova ejecta,
but the interpretation has not been settled.
On the other hand, the mass and velocity of the kilonova ejecta
are well constrained by the UV/optical/IR observations.
We simulate the nonthermal emission emitted from
electrons accelerated at the shock propagating in the CBM.
If the ambient density is higher than $\sim 10^{-2}~\mbox{cm}^{-3}$,
we can expect detections of radio, optical/IR, and X-ray
counterparts in a few years.
If the early X-ray and radio counterparts are emitted from
the shock produced by the additional low-mass ejecta of $v \sim 0.5 c$--$0.8 c$
\citep[see also][]{moo17},
such a high-density case can be justified.
In this case, the off-axis GRB model with a low CBM density should be reconsidered.
Even for $n \sim 10^{-3}~\mbox{cm}^{-3}$, however, optimistic parameter sets
with a low $\eta$ lead to a detectable radio flux.

\citet{mur17} proposed another type of the
delayed electromagnetic counterparts due to the long-lasting activity
of the central objects, such as the disk-driven outflows or pulsar winds.
The emission timescale in the long-lasting engine model
is shorter; typically the flux peaks in a year,
differently from the case in the kilonova ejecta model.
Future follow-up observations are important to verify such various
possibilities.

\acknowledgements

First, we are grateful to the anonymous referee for helpful advise.
This work is supported by the joint research program of
the Institute for Cosmic Ray Research (ICRR), the University of Tokyo,
and Grants-in-Aid for Scientific
Research nos. 15K05069 and 16K05291 (K.A.) from the Ministry
of Education, Culture, Sports, Science and Technology
(MEXT) of Japan.
We also appreciate S. Kisaka for his useful comments.


\begin{thebibliography}{}
\bibitem[Abbott et al. (2017a)]{abb17b}
Abbott, B. P., Abbott, R., Abbott, T. D., et al., 2017a, \apj, 848, L12
\bibitem[Abbott et al. (2017b)]{abb17c}
Abbott, B. P., Abbott, R., Abbott, T. D., et al., 2017b, \apj, 848, L13
\bibitem[Abbott et al. (2017c)]{abb17a}
Abbott, B. P., Abbott, R., Abbott, T. D., et al., 2017c, \prl, 119, 161101
\bibitem[Alexander et al. (2017)]{ale17}
Alexander, K. D., Berger, E., Fong, W., et al., 2017, \apj, 848, L21
\bibitem[Arcavi et al. (2017)]{arc17}
Arcavi,	I., Hosseinzadeh, G., Howell, D. A., et al., 2017, \nat, 551, 64
\bibitem[Bromberg et al. (2017)]{bro17}
Bromberg, O., Tchekhovskoy, A., Gottlieb, O., Nakar, E., \& Piran, T., 2017, arXiv:1710.05897
\bibitem[Chandra \& Frail (2012)]{cha12}
Chandra, P., \& Frail, D. A., 2012, \apj, 746, 156
\bibitem[Coulter et al. (2017)]{cou17}
Coulter, D. A., Foley, R. J., Kilpatrick, C. D., et al., 2017, Sci, 358, 1556
\bibitem[Cowperthwaite et al. (2017)]{cow17}
Cowperthwaite, P. S., Berger, E., Villar, V. A., et al., 2017, \apj, 848, L17
\bibitem[Fukushima et al. (2017)]{fuk17}
Fukushima, T., To, S., Asano, K., \& Fujita, Y., 2017, \apj, 844, 92
\bibitem[Gottlieb et al. (2017)]{got17}
Gottlieb, O., Nakar, E., Piran, T., \& Hotokezaka, K., 2017, arXiv:1710.05896
\bibitem[Hotokezaka et al. (2013)]{hot13}
Hotokezaka, K., Kiuchi, K., Kyutoku, K., et al., 2013, \prd, 87, 024001
\bibitem[Hotokezaka et al. (2016)]{hot16}
Hotokezaka, K., Nissanke, S., Hallinan, G., et al., 2016, \apj, 831, 190
\bibitem[Hotokezaka \& Piran (2015)]{hot15}
Hotokezaka, K., \& Piran, T., 2015, \mnras, 450, 1430
\bibitem[Im et al. (2017)]{im17}
Im, M., Yoon, Y., Lee, S.-K. J., et al., 2017, \apj, 849, L16
\bibitem[Ioka \& Nakamura (2017)]{iok17}
Ioka, K., \& Nakamura, T., 2017, arXiv:1710.05905
\bibitem[Kasliwal et al. (2017)]{kas17}
Kasliwal, M. M., Nakar, E., Singer, L. P., et al., 2017, Sci, 358, 1559
\bibitem[Kyutoku et al. (2014)]{kyu14}
Kyutoku, K., Ioka, K., \& Shibata, M., 2014, \mnras, 437, L6
\bibitem[Lattimer \& Schramm (1974)]{lat74}
Lattimer, J. M., \& Schramm, D. N., 1974, \apj, 192, L145
\bibitem[Lazzati et al. (2017)]{laz17}
Lazzati, D., Perna, R., Morsony, B. J., L\'opez-C\'amara, D.,
Cantiello, M., Ciolfi, R., Giacomazzo, B., \& Workman, J. C., 2017, arXiv:1712.03237
\bibitem[Li \& Paczy\'nski (1998)]{li98}
Li, L.-X., \& Paczy\'nski, B., 1998, \apj 507, L59
\bibitem[Margutti et al. (2017)]{mar17}
Margutti, R., Berger, E., Fong, W., et al., 2017, \apj, 848, L20
\bibitem[Metzger et al. (2010)]{met10}
Metzger, B. D., Mart\'inez-Pinedo, G., Darbha, S., et al., 2010, \mnras, 406, 2650
\bibitem[Mooley et al. (2017)]{moo17}
Mooley, K. P., Nakar, E., Hotokezaka, K., et al., 2017, arXiv:1711.11573
\bibitem[Murguia-Berthier et al. (2017)]{murg17}
Murguia-Berthier, A., Ramirez-Ruiz, E., Kilpatrick, C. D., et al., 2017, \apj, 848, L34
\bibitem[Murase et al. (2017)]{mur17}
Murase, K., Toomey, M. W., Fang, K., et al., 2017, 	arXiv:1710.10757
\bibitem[Nakar \& Piran (2011)]{nak11}
Nakar, E., \& Piran, T., 2011, \nat, 478, 82
\bibitem[Pennanen et al. (2014)]{pen14}
Pennanen, T., Vurm, I., \& Poutanen, J. 2014, \aap, 564, A77
\bibitem[Petropoulou \& Mastichiadis (2009)]{pet09}
Petropoulou, M., \& Mastichiadis A. 2009, \aap, 507, 599
\bibitem[Piran et al. (2013)]{pir13}
Piran, T., Nakar, E., \& Rosswog, S., 2013, \mnras, 430, 2121
\bibitem[Rosswog et al. (1999)]{ros99}
Rosswog, S., Liebend\"orfer, M., Thielemann, F.-K., Davies, M. B., Benz, W.,
\& Piran, T., 1999, \aap, 341, 499
\bibitem[Ruan et al. (2017)]{rua17}
Ruan, J. J., Nynka, M., Haggard, D., Kalogera, V., \& Evans, P., 2017, arXiv:1712.02809
\bibitem[Rosswog et al. (2013)]{ros13}
Rosswog, S., Piran, T., \& Nakar, E., 2013, \mnras, 430, 2585
\bibitem[Smartt et al. (2017)]{sma17}
Smartt, S. J., Chen, T.-W., Jerkstrand, A., et al., 2017, \nat, 551, 75
\bibitem[Takami et al. (2014)]{tak14}
Takami, H., Kyutoku, K., \& Ioka, K., 2014, \prd, 89, 063006
\bibitem[Tanaka \& Hotokezaka (2013)]{tan13}
Tanaka, M., \& Hotokezaka, K., 2013, \apj, 775, 113
\bibitem[Tanaka et al. (2017)]{tan17}
Tanaka, M., Utsumi, Y., Mazzali, P. A., et al., 2017, PASJ, 69, 102
\bibitem[Troja et al. (2017)]{tro17}
Troja, E., Piro, L., van Eerten, H., et al., 2017, \nat, 551, 71
\bibitem[Uhm \& Zhang (2014)]{uhm14}
Uhm, Z. L., \& Zhang, B. 2014, \apj, 780, 82
\bibitem[Valenti et al. (2017)]{val17}
Valenti, S., Sand, D. J., Yang, S., et al., 2017, \apj, 848, L24
\end{thebibliography}
\end{document}